\documentclass[copyright,creativecommons]{eptcs}

\providecommand{\event}{}
\providecommand{\volume}{??}
\providecommand{\anno}{20??}
\providecommand{\firstpage}{1}
\usepackage{breakurl}

\def\titlerunning{Random CSPs}
\def\authorrunning{Amin Coja-Oghlan}

\def\cE{{\cal E}}

\newcommand\eul{\mathrm{e}} 
\newcommand\eps{\varepsilon} 
\newcommand\Erw{\mathrm{E}} 
\newcommand\pr{\mathrm{P}}

\newcommand\brk[1]{\left\lbrack{#1}\right\rbrack} 
\newcommand{\whp}{w.h.p.}

\begin{document} 

\title{Random Constraint Satisfaction Problems}
 
\author{Amin Coja-Oghlan\thanks{Supported by EPSRC grant EP/G039070/1.}
\institute{University of Edinburgh, School of Informatics, Edinburgh EH8 9AB, UK \email{acoghlan@inf.ed.ac.uk}}}

\maketitle 

\begin{abstract}
Random instances of constraint satisfaction problems such as $k$-SAT provide challenging benchmarks.
If there are $m$ constraints over $n$ variables there is typically a large range of densities $r=m/n$
where solutions are known to exist with probability close to one due to non-constructive arguments.
However, no algorithms are known to find solutions efficiently with a non-vanishing probability at even much lower densities.
This fact appears to be related to a phase transition in the set of all solutions.
The goal of this extended abstract is to provide a perspective on this phenomenon, and on the computational challenge
that it poses.
\end{abstract}

\section{A computational challenge}

Numerous \emph{constraint satisfaction problems} (``CSPs'') are well known to be NP-hard.
Examples of such problems are for any $k\geq3$
\begin{description}
\item[$k$-SAT.] 
	The input is a propositional formula in conjunctive normal form
		$\Phi=\Phi_1\wedge\cdots\wedge\Phi_m$,
	where each clause $\Phi_i$ is a disjunction of $k$ literals over a set of Boolean variables
		$\{x_1,\ldots,x_n\}$.
	The goal is to decide whether there is an assignment of  $x_1,\ldots,x_n$
	such that the expression $\Phi$ evaluates to true, and if so, to find such a satisfying assignment.
\item[$k$-NAE.]
	The input is  a propositional formula as in $k$-SAT.
	The objective is to decide whether there is an assignment of $x_1,\ldots,x_n$
	such that each clause contains both a true and a false literal (``Not All Equal'').
\item[$k$-coloring.]
	Given a (simple, undirected) graph $G=(V,E)$, decide whether
	there is a \emph{$k$-coloring}, i.e., an assignment $V\rightarrow\{1,\ldots,k\}$ of
	``colors'' to the vertices such that for any edge $e=\{v,w\}\in E$ the vertices
	$v,w$ are assigned different colors.
\end{description}
We will call the desired type of assignment in each case a \emph{solution}.

Since the above problems are NP-hard, no efficient algorithm
is known to solve \emph{all} possible problem instances.
However, the theory of NP-hardness does not provide us with an explicit class
of ``hard'' problem instances.
It just shows that all NP-complete problems are ``equally hard''.
Furthermore, NP-hardness merely suggests that there \emph{exist} hard problem instances.
This does not rule out efficient heuristic algorithms that succeed
on ``most'' inputs.

Yet it is surprisingly simple to generate problem instances that seem to elude all known heuristics.
In fact, in all of the above problems, \emph{randomly} generated instances provide extremely challenging inputs.
Throughout we denote the number of \emph{variables}  by $n$ and the number of \emph{constraints} by $m$.
In the case of $k$-SAT and $k$-NAE ``variable'' has the obvious meaning.
In $k$-coloring ``variables'' refers to the vertices of the graph.
Moreover, the constraints are the edges (in $k$-coloring) resp.\ the clauses (in $k$-SAT and $k$-NAE).
We consider random problem instances generated by just picking a set of $m$ constraints uniformly at random.
In $k$-SAT or $k$-NAE this yields a random propositional formula $F_k(n,m)$,
and in $k$-coloring a random graph $G(n,m)$.
Generally we will be interested in ``large'' instances, i.e., $n\rightarrow\infty$.
Furthermore, mostly the \emph{constraint density} $r=m/n$ will remain bounded as $n$ gets large.
We say that the random instance has a property $\cE$ \emph{with high probability} (``\whp'')
if the probability that $\cE$ holds tends to one as $n\rightarrow\infty$.

For a large range of densities $r$ non-constructive arguments show that
the random instance $F_k(n,m)$ or $G(n,m)$ has a solution \whp,
but no efficient algorithm is known to find one with a non-vanishing probability.
Thus, random constraint satisfaction problems pose an algorithmic challenge.
They have withstood more than 20 years of extensive research efforts.
The aim of this short paper is to provide a perspective on this
instructive class of instances.

\section{For what densities do solutions exist?}

In each of $k$-SAT, $k$-NAE, and $k$-coloring there is a \emph{sharp threshold} $r_k$ such that
for densities $r<r_k-\eps$ there is a solution \whp, while for densities $r>r_k+\eps$ no solution exists \whp,
for any fixed $\eps>0$~\cite{AchFried,EhudHunting}.
Actually the threshold $r_k$ is \emph{non-uniform}, i.e., $r_k=r_k(n)$ is not a fixed number but may depend on $n$.
However, $r_k(n)$ is conjectured to converge.

The threshold $r_k=r_k(n)$ is not known precisely for any $k\geq3$ (and large $n$), but asymptotically tight bounds in the large $k$ limit are.
An upper bound on $r_k$ can be obtained by computing the \emph{expected} number of solutions.
If $r$ is such that the expected number of solutions is $o(1)$ as $n\rightarrow\infty$, then $r_k\leq r$ by Markov's inequality.
Proofs of this kind are called \emph{first moment arguments} (cf.~\cite{AlSp}).
They show that $r_k\leq 2^k\ln2$ in $k$-SAT, $r_k\leq 2^{k-1}\ln2$ in $k$-NAE, and $r_k\leq k\ln k$ in $k$-coloring.

For general values of $k$ the best current lower bounds on $r_k$ are obtained via the \emph{second moment method}.
The simple idea is to bound the expectation of the squared number of solutions for a given density $r$.
To be precisely, let $X=X(n,m)$ denote the (random) number of solutions.
If $\Erw(X)\gg1$ and $\Erw(X^2)=O(\Erw(X)^2)$, then the \emph{Paley-Zigmund inequality}
	$$\pr\brk{X>0}\geq\frac{\Erw(X)^2}{\Erw(X^2)}$$
entails that the probability that there is a solution remains bounded away from zero for arbitrarily large $n$.
Hence, the sharp threshold result implies that $r_k\geq r=m/n$.

The second moment method can be applied quite directly to both $k$-NAE and $k$-coloring~\cite{nae,kcol}.
In the latter case computing the second moment amounts to a challenging optimization problem over the Birkhoff polytope.
In $k$-SAT, by contrast, it is necessary to assign certain weights to the solutions~\cite{yuval}.
The result is that
	$r_k\geq2^k\ln2-O(k)$ in $k$-SAT,
	$r_k\geq2^k\ln2-O(1)$ in $k$-NAE,
	and $r_k\geq k\ln k-O(\ln k)$ in $k$-coloring.
Thus, asymptotically for large $k$ these bounds match the first moment upper bounds, up to second order terms.

\section{Finding solutions efficiently}

In all three problems the density up to which solutions are known to exist exceeds the density up to which
efficient algorithms are known to find any significantly.
In $k$-SAT (resp.~$k$-NAE) no algorithm is known to find solutions beyond $r=2^k\ln(k)/k$ (resp.\ $r=2^{k-1}\ln(k)/k$)
for large $k$.
This is by a factor $k/\ln(k)$ below the threshold density.
Moreover, no algorithm is known to find $k$-colorings of $G(n,m)$ for $r=m/n>\frac12k\ln k$ for large $k$ (a factor of $2$).
This is in spite of intensive research on the subject.
In the sequel we follow the discussion in~\cite[Section~1.1]{AchACORicci}.

To describe the class of algorithms that have been suggested and/or analyzed we need the notion of a \emph{factor graph}.
This is a bipartite graph associated with a problem instance.
Its vertices are the constraints and the variables.
Each variable is connected with all the constraints that it occurs in.
In the case of random CSPs with $r$ bounded as $n\rightarrow\infty$ the factor graph
has girth $\Omega(\ln n)$ \whp\ (after the removal of $O(1)$ constraints).
Hence, for each node $v$ the subgraph $N_\omega(v)$ spanned by the vertices $w$ that are at distance
at most $\omega$ from $v$ is a tree \whp, provided that $\omega\ll\ln n$.

Most algorithms that have been suggested are \emph{local}.
That is, the value that the algorithm assigns to a variable $x$
only depends on the constraints and variables in an $\omega$-neighborhood $N_\omega(x)$
for some fixed number $\omega$ (i.e., independent of $n$).
In fact, mostly $\omega=1$ or $\omega=2$.
Furthermore, most algorithms do not backtrack.
That is, once a variable has been assigned, its value will never change.

The \texttt{UnitClause} algorithm for $k$-SAT is a prototypical example.
Initially the algorithm considers all variables unassigned.
In each step it selects a variable and assigns it for good.
In step $t$ the algorithm checks if there is a \emph{unit clause},
i.e., a clause in which precisely $k-1$ literals are false due to previous assignments.
If so, it selects an unassigned variable $x_t$ that occurs in a unit clause
and sets $x_t$ so as to satisfy the unit clause.
If not, the algorithm selects a variable $x_t$ randomly and assigns it a random value.
In the limit of large $k$ this simple linear time algorithm finds a satisfying assignment
with a non-vanishing probability for densities up to $r\sim\frac{\eul}2\cdot\frac{2^k}k$~\cite{ChaoFranco2}.
The best current algorithm for random $k$-SAT is local as well
(with $\omega=3$) and succeeds up to $(1-\eps_k)2^k\ln(k)/k$,
where $\eps_k\rightarrow0$~\cite{BetterAlg}.

In graph coloring the situation is similar.
A very simple greedy algorithm ($\omega=2$) succeeds up to density $r=(\frac12-\eps_k)k\ln k$ for large $k$.
A slightly better local algorithm ($\omega=2$ as well) actually works up to $r=\frac12k\ln k$~\cite{focs}.

Given the simplicity of these algorithms, their (rigorous) analyses can be surprisingly demanding.
They are mostly based on tracking the execution of the algorithm by either differential equations,
Markov chains, or martingales.
The use of these techniques seems limited to algorithms with small depth, say $\omega=2$ or $\omega=3$.
Furthermore, these methods do not seem sufficient for analyzing algorithms that reassign variables.

A class local of algorithms with larger depth $\omega$ have been put forward on the basis of ideas from the
statistical mechanics of disordered systems~\cite{BMZ}.
Here $\omega$ is independent of $n$ but not bounded \emph{a priori}.
In other words, it has to be chosen sufficiently large in terms of $r$ and $k$.
In each round the algorithm aims to assign one variable (for good).
To this end the algorithm performs for each variable $x$ a computation that depends on the subgraph
$N_\omega(x)$ and the values that have been assigned to the variables in that subgraph previously.
For instance, in $k$-SAT the algorithm considers the sub-formula of the input $k$-SAT formula
that corresponds to $N_\omega(x)$.
It computes the probability that in a \emph{random} satisfying assignment of that subformula,
given the values of all previously assigned variables in it, $x$ takes the value true/false.
Then, the algorithm selects the variable for which this computation yields the largest bias
towards either value and assigns the variable accordingly.
The computation on the sub-instance $N_\omega(x)$ can be performed efficiently,
because $N_\omega(x)$ is acyclic \whp\
In fact, the computation can be implemented to run simultaneously for all variables
by means of a message passing procedure (``Belief Propagation'').
The Survey Propagation algorithm is a somewhat more involved variant of this strategy
(see~\cite{BMZ} for details).

Since this type of algorithm crucially requires large $\omega$ (for the estimates of the
marginals to be accurate), its rigorous analysis is beyond current methods.
Experimentally algorithms based on this scheme, namely, Belief/Survey Propagation guided decimation,
outperform any other known ones by far for small $k$.
However, for large $k$ experimental evidence is difficult to come by.
For instance, in $k$-SAT the relevant density scales exponentially in~$k$.

\section{An algorithmic barrier?}

On the basis of non-rigorous but sophisticated techniques from statistical mechanics a hypothesis has been put
forward that might explain the demise of local algorithms way below the threshold for the existence of solutions~\cite{pnas}.
This hypothesis concerns the \emph{solution space}.
For a CSP instance $\Phi$ we let $S(\Phi)$ signify the set of all solutions of $\Phi$.
For instance, if $\Phi$ is a $k$-SAT formula with $n$ variables, then $S(\Phi)\subset\{0,1\}^n$ is the set of all satisfying assignments.
Similarly, if $\Phi$ is a graph on $n$ vertices, then $S(\Phi)\subset\{1,\ldots,k\}^n$ is the set of all $k$-colorings.
We turn the set $S(\Phi)$ into a graph by considering $\sigma,\tau\in S(\Phi)$ adjacent if
their Hamming distance equals one.

For random CSP instances with densities below the threshold $r_k$ the size of $S(\Phi)$ is exponential in $n$ \whp\
More precisely, in both $k$-coloring and $k$-NAE we have $|S(\Phi)|=\Erw(|S(\Phi)|)\cdot\exp(o(n))$ \whp,
and the first moment $\Erw(|S(\Phi)|)$ is easily computed~\cite{AchACO}.
By contrast, in $k$-SAT we have $|S(\Phi)|\leq\Erw(|S(\Phi)|)\cdot\exp(-\Omega(n))$ \whp,
but $|S(\Phi)|\geq\Erw(|S(\Phi)|)\cdot\exp(-\zeta_k n)$ \whp, where $\zeta_k\rightarrow0$ exponentially for large $k$~\cite{AchACO,AchACORicci}.

The \emph{dynamic replica symmetry breaking} (``dRSB'') hypothesis states that in each of $k$-SAT, $k$-NAE, and $k$-coloring
there is a density $r_{dRSB}<r_k$ below the threshold for the existence of solutions where the shape of the set $S(\Phi)$ undergoes
a phase transition.
Furthermore, $r_{dRSB}$ coincides asymptotically with the density up to which local algorithms are known to find solutions.
That is, in the large $k$ limit
$r_{dRSB}\sim 2^k\ln(k)/k$ in $k$-SAT, $r_{dRSB}\sim 2^{k-1}\ln(k)/k$ in $k$-NAE, and $r_{dRSB}\sim\frac12k\ln k$ in $k$-coloring.
According to the dRSB hypothesis, for densities $r<r_{dRSB}$ the graph $S(\Phi)$ is essentially connected \whp\
More precisely, there is a single component that contains a $1-o(1)$ fraction of all solutions.
By contrast, for densities $r>r_{dRSB}$ there are exponentially many components, none of which contains
more than an exponentially small fraction of all solutions.
This means that for $r<r_{dRSB}$ the correlations among the variables that shape the set $S(\Phi)$ are purely local,
whereas for $r>r_{dRSB}$ long range correlations arise.

Confirming and elaborating on this hypothesis, we recently established a good bit of the dRSB phenomenon rigorously~\cite{AchACO}.
We proved that beyond the conjectured densities $r_{dRSB}$ the set $S(\Phi)$ decomposes into exponentially small well-separated components
in $k$-SAT, $k$-NAE, and $k$-coloring.
Furthermore, each component is very rigid locally.
To be precise, we say that a variable $x$ is
\emph{frozen} in a solution $\sigma$ if any solution $\tau$ such that $\tau(x)\not=\sigma(x)$
has at least a linear Hamming distance $\Omega(n)$ from $\sigma$.
In other words, changing the value of $x$ necessitates changing the values of $\Omega(n)$ other variables.
Then for $r>(1+\eps_k)r_{dRSB}$ in all but a $o(1)$-fraction of all solutions all but an $\eps_k$-fraction of the variables are frozen \whp,
where $\eps_k\rightarrow0$ for large $k$.

This suggests that on random instances with density $r>(1+\eps_k)r_{dRSB}$ local algorithms are unlikely to succeed.
For a local search algorithm assigns variables $x$ only on the basis of
the values of variables that have distance at most $\omega$ from $x$ in the factor graph,
where $\omega=O(1)$ is bounded as $n\rightarrow\infty$.
But the presence of frozen variables
yields mutual constraints on the values that can be assigned to variables at distance $\Omega(\ln n)$ from $x$ in the factor graph.
Local algorithms do not seem capable of catching these long-range effects.

The above discussion applies to ``large'' values of $k$ (say, $k\geq 10$). 
Non-rigorous arguments as well as experimental evidence~\cite{Lenka} suggest that
the picture is quite different and rather more complicated for ``small'' $k$.
This may be the reason why local algorithms such as Survey Propagation guided decimation fare extremely well
for, e.g., random $k$-SAT with $k=3,4,5$.
Whether or not algorithms of this type
 succeed beyond $(1+\eps)r_{dRSB}$ for large $k$ and any fixed $\eps>0$ remains an important open problem.
A plausible scenario may be that such algorithms succeed up to $r=(1+\eps_k)r_{dRSB}$ for some $\eps_k\rightarrow0$.

\section{Conclusion}

Random instances of constraint satisfaction problems exhibit a phase transition with respect to the existence of solutions.
In addition, there is strong evidence that at a much lower constraint density a further transition takes place that affects
the performance of local algorithms.
In statistical mechanics terms, this is know as dynamic replica symmetric breaking.
Roughly speaking, while below the density $r_{dRSB}$ conceptually fairly simple algorithms find solutions efficiently,
no efficient algorithm is known to find any beyond that density (for general values of $k$).
This appears to be due to a transition in the geometry of the set of all solutions,
which shatters into exponentially small components and exhibits frozen variables  beyond $r_{dRSB}$.
Coping with problem instances of this type poses an algorithmic challenge.

Virtually all algorithms that have been suggested/analyzed for sparse random CSPs are (essentially) local.
It seems plausible that such algorithms have a hard time catching the long-range correlations that occur beyond $r_{dRSB}$.
However, proving this in any generality is an open problem.

Global algorithmic techniques such as spectral methods or semidefinite programming apply to classes of randomly generated
CSP instances that have essentially a single solution and a sufficiently high constraint density (way beyond the threshold $r_k$
for the existence of solutions in the models discussed here)~\cite{AlonKahale,FlaxmanSpec}.
One way of generating such instances is by ``planting'' a solution in an otherwise random instance.
The success of spectral methods implies the success of Belief Propagation in, for instance, random 3-coloring~\cite{BPSpec}.
But global methods do not seem to apply to instances of relatively low density (i.e., below the threshold $r_k$).

Thus, no efficient algorithms are known to solve random CSP instances with density $r_{dRSB}<r<r_k$
for general $k$.
Moreover, this seems to be a fairly universal fact, independent of the precise CSP under consideration.
It might be interesting to  investigate how alternative models of computation or alternative algorithmic paradigms
fare on such inputs.

\end{document}